\documentclass[12pt,onecolumn]{revtex4}

\begin{document}

\title{Charged Rotating Black Branes in anti-de Sitter Einstein-Gauss-Bonnet Gravity}
\author{M. H. Dehghani}\email{dehghani@physics.susc.ac.ir}

\address{Physics Department and Biruni  Observatory,
         Shiraz University, Shiraz 71454, Iran\\ and\\
         Institute for Studies in Theoretical Physics and Mathematics (IPM)\\P.O. Box 19395-5531, Tehran, Iran}

\begin{abstract}
We present a new class of charged rotating solutions in the
Einstein-Gauss-Bonnet gravity with a negative cosmological
constant. These solutions may be interpreted as black brane
solutions with two inner and outer event horizons or an extreme
black brane depending on the value of the mass parameter $m$. We
also find that the Killing vectors are the null generators of the
event horizon. The physical properties of the brane such as the
temperature, the angular velocity, the entropy, the electric
charge and potential are computed. We also compute the action and
the Gibbs potential as a function of temperature and angular
velocity for the uncharged solutions, and compute the angular
momentum and the mass of the black brane through the use of Gibbs
potential. We show that these thermodynamic quantities satisfy the
first law of thermodynamics. We also perform a local stability
analysis of the asymptotically AdS uncharged rotating black brane
in various dimensions and show that they are locally stable for
the whole phase space both in the canonical and grand-canonical
ensemble. We found that the thermodynamic properties of
Gauss-Bonnet rotating black branes are completely the same as
those without the Gauss-Bonnet term, although the two solutions
are quite different.
\end{abstract}

\maketitle
\section{Introduction}

It is a general belief that Einstein's gravity is a low-energy
limit of a quantum theory of gravity which is still unknown.
Irrespective of the fundamental nature of quantum gravity, there
should be a low-energy effective action which describes gravity at
the classical level. This effective action will consist the
Einstein-Hilbert action plus higher derivative (HD) terms. The
appearance of HD gravitational terms can be seen, for example, in
the renormalization of quantum field theory in curved spacetime
\cite{BDav}, or in the construction of low-energy effective action
of string theory \cite{Wit1}.

Among the theories of gravity with higher derivative curvature
terms, the Gauss-Bonnet gravity is of some special features in
some sense. Indeed, in order to have a ghost-free action, the
quadratic curvature corrections to the Einstein-Hilbert action
should be proportional to the Gauss-Bonnet term \cite{Zw,Des}.
This combination also appear naturally in the next-to-leading
order term of the heterotic string effective action, and plays a
fundamental role in Chern-Simons gravitational theories
\cite{Cham}. From a geometric point of view, the combination of
the Einstein-Gauss-Bonnet terms constitutes, for five dimensional
spacetimes, the most general Lagrangian producing second order
field equations, as in the four-dimensional gravity which the
Einstein-Hilbert action is the most general Lagrangian producing
second order field equations \cite{Lov}. These facts provide a
strong motivation for considering new exact solutions of the
Einstein-Gauss-Bonnet gravity. Indeed, it is interesting to
explore black holes in generalized gravity theories in order to
discover which properties are peculiar to Einstein's gravity, and
which are robust features of all generally covariant theories of
gravity.

Due to the nonlinearity of the field equations, it is very
difficult to find out nontrivial exact analytical solutions of
Einstein's equation with the higher derivative terms. In most
cases, one has to adopt some approximation methods or find
solutions numerically. However, static spherically symmetric black
hole solutions of the Gauss-Bonnet gravity have been found in Ref.
\cite{Des}. Black hole solutions with nontrivial topology in this
theory have been also studied in Refs. \cite{Cai,Aros,Ish}. The
thermodynamics of charged static spherically symmetric black hole
solutions has been considered in \cite{Odin}. All of these known
solutions are static. In this paper we will find the non-static
asymptotically anti-de Sitter charged rotating black brane
solutions in the Einstein-Gauss-Bonnet gravity.

The outline of our paper is as follows. We present a new class of
charged rotating solutions with all rotation parameters for the
Einstein-Gauss-Bonnet-Maxwell equation in Sec. \ref{Sol}.
In Sec. \ref{Prop} we consider the properties of these solutions in $%
(n+1)$ dimensions, and obtain their entropy, temperature, charge
and electric potential. In Sec. \ref{Act}, we compute the action,
the angular momentum and the mass of the black brane by obtaining
the Gibbs potential of the system. Section \ref{Stab} will be
devoted to the local stability analysis of the black brane in both
the canonical and the grand-canonical ensemble. We finish our
paper with some concluding remarks.


\section{Charged rotating solutions of Einstein-Gauss-Bonnet's equation \label
{Sol}}

The gravitational action of an $(n+1)$-dimensional asymptotically
anti-de Sitter spacetimes with the Gauss-Bonnet term in the
presence of an electromagnetic field can be written as
\begin{equation}
I=-\frac 1{16\pi }\int
d^{n+1}x\sqrt{-g}\{R+\frac{n(n-1)}{l^2}+\alpha (R_{\mu \nu \gamma
\delta }R^{\mu \nu \gamma \delta }-4R_{\mu \nu }R^{\mu \nu
}+R^2)+F_{\mu \nu } F^{\mu \nu }\}, \label{Actg}
\end{equation}
where $F_{\mu \nu }=\partial _\mu A_\nu -\partial _\nu A_\mu $ is
the electromagnetic tensor field, and $A_\mu $ is the vector
potential. $\alpha $ is the Gauss-Bonnet coefficient with
dimension $(length)^2$ and is positive in the heterotic string
theory \cite{Des}. So we restrict ourselves to the case $\alpha
\ge 0$. The first two terms are the Einstein-Hilbert action with
negative cosmological constant which are the leading ones while
the Gauss-Bonnet term is the next-to-leading term in low-energy
stringy effective action. Varying the action over the metric
tensor $g_{\mu \nu }$ and electromagnetic field $F_{\mu \nu }$,
the equations of gravitational and electromagnetic fields are
obtained as
\begin{eqnarray}
&&\ R_{\mu \nu }-\frac 12g_{\mu \nu }R+\frac{n(n-1)}{2l^2}g_{\mu
\nu }-\alpha \{\frac 12g_{\mu \nu }(R_{\gamma \delta \lambda
\sigma }R^{\gamma \delta \lambda \sigma }-4R_{\gamma \delta
}R^{\gamma \delta }+R^2)  \nonumber
\\
&&-2RR_{\mu \nu }+4R_{\mu \gamma }R_{\ \nu }^\gamma +4R_{\gamma
\delta }R_{\mu \nu }^{\gamma \ \delta }-2R_{\mu \gamma \delta
\lambda }R_\nu ^{\ \gamma \delta \lambda }\}=T_{\mu \nu },
\label{Geq}
\end{eqnarray}
\begin{equation}
\nabla _\mu F_{\mu \nu }=0, \label{EMeq}
\end{equation}
where $T_{\mu \nu}$ is the electromagnetic stress tensor
\begin{equation}
T_{\mu \nu}=2 {F^\lambda \ _\mu} F_{\lambda \nu}
-\frac{1}{2}F_{\lambda \sigma } F^{\lambda \sigma}g_{\mu \nu}.
\label{Str}
\end{equation}
Equation (\ref{Geq}) does not contain the derivative of the
curvatures, and therefore the derivatives of the metric higher
than two do not appear. Thus, the Gauss-Bonnet gravity is a
special case of higher derivative gravity. In this section we want
to obtain the rotating solutions of Eqs.(\ref{Geq})-(\ref{Str}).
We first consider the metric with one rotation parameter.

\subsection{Metric with one rotation parameter}

We assume the metric and the vector potential have the following forms
\begin{eqnarray}
ds^2 &=&-f(r)(\Xi dt-ad\phi )^2+\frac{r^2}{l^4}(adt-\Xi l^2d\phi )^2+\frac{%
dr^2}{f(r)}+r^2d\Omega ^2,  \label{met1a} \\
\Xi ^2 &=&1+\frac{a^2}{l^2},  \label{met1b} \\
A_\mu &=&qh(r)(\Xi \delta _\mu ^0-a\delta _\mu ^2),  \label{met1c}
\end{eqnarray}
where $d\Omega ^2$ is the Euclidean metric on the
$(n-2)$-dimensional submanifold with volume $\omega _{n-2}$. The
functions $f(r)$ and $h(r)$ may be obtained by solving the field
equations (\ref{Geq}) and (\ref{EMeq}). This form of the metric
was first introduced by Lemos \cite{Lem} in four dimensions, and
it has been
generalized by Awad \cite{Awad} for higher dimensions. Using Eq. (\ref{EMeq}%
) one can show that $h(r)=C_1/r^{n-2}$, where $C_1$ is an arbitrary real
constant. To find the function $f(r)$, one may use any components of Eq. (%
\ref{Geq}). The simplest equation is the $rr$-component of these
equations which can be written as
\begin{equation}
r^5f^{^{\prime }}+2r^4f-4r^6+2q^2-4\alpha r^3f(r)f^{^{\prime }}=0,
\label{rrcom}
\end{equation}
where the prime denotes a derivative with respect to the $r$
coordinate. The solution of Eq. (\ref{rrcom}) is
\begin{equation}
f(r)=\frac{r^2}{2(n-2)(n-3)\alpha }\left\{1\pm
\left[1-4(n-2)(n-3)\alpha
\left( \frac 1{l^2}+\frac{(2n-4)C_1^2}{(n-1)r^{2n-2}}-\frac{C_2}{(n-1)r^n}%
\right) \right]^{1/2} \right\} ,  \label{Fg}
\end{equation}
where $C_2$ is an arbitrary constant. As one can see from Eq.
(\ref{Fg}), the solution has two branches with ''$-$'' and ''$+$''
signs. In order to have an asymptotically anti-de Sitter
spacetime, we
should choose the branch with ''$-$'' sign. Indeed if one chooses the ''$-$%
'' sign, then the asymptotic behavior of the Ricci scalar for
$\alpha =0$ is $-n(n+1)/l^2$, which is the case for asymptotic AdS
spacetimes, while the Ricci scalar for ''$+$'' sign branch
diverges as $\alpha $ goes to
zero. Choosing $C_1^2=(n-1)/(2n-4)q^2$ and $C_2=(n-1)m$, then in the case of $%
\alpha =0$, one obtains the function $f(r)$ of Ref. \cite{Awad} as
expected. The thermodynamics and the no-hair theorem of the latter
case in four dimension have been studied in Ref. \cite{Deh1,Deh2}.

When $m$ and $q$ are zero, the vacuum solution is
\begin{equation}
f(r)=\frac{r^2}{2(n-2)(n-3)\alpha }\left( 1-\sqrt{1-\alpha \frac{4(n-2)(n-3)%
}{l^2}}\right) .  \label{Fg0}
\end{equation}
Equation (\ref{Fg0}) shows that for a positive value of $\alpha $,
this parameter should be less than $\alpha \leq l^2/4(n-2)(n-3)$.
Also note that the AdS solution of the theory has the effective
cosmological constant
\[
l_{\mathrm{eff}}^2=2(n-2)(n-3)\alpha \left( 1-\sqrt{1-\frac{%
4(n-2)(n-3)\alpha }{l^2}}\right)^{-1} .
\]

\subsection{The generalized metric with all rotation parameters}

The rotation group in $(n+1)$-dimensions is $SO(n)$ and therefore
the number of independent rotation parameters for a localized
object is equal to the number of Casimir operators, which is
$[n/2]\equiv k$, where $[n/2]$ is the integer part of $n/2$. The
generalization of the metric (\ref{met1a}) with all rotation
parameters is \cite{Awad}
\begin{eqnarray}
ds^2 &=&-f(r)\left( \Xi dt-{{\sum_{i=1}^k}}a_id\phi _i\right) ^2+\frac{r^2}{%
l^4}{{\sum_{i=1}^k}}\left( a_idt-\Xi l^2d\phi _i\right) ^2\nonumber \\
&&-\frac{r^2}{l^2}{%
\sum_{i=1}^k}(a_id\phi _j-a_jd\phi _i)^2+\frac{dr^2}{f(r)}+r^2d\Omega ^2,
\nonumber \\
\Xi ^2 &=&{{\sum_{i=1}^k}}\left( 1+\frac{a_i^2}{l^2}\right) ,
\label{met2}
\end{eqnarray}
where $d\Omega ^2$ is now the Euclidean metric on the
$(n-k-1)$-dimensional submanifold with volume $\omega _{n-k-1}$.
Again one can show the asymptotically AdS solutions of Eqs.
(\ref{Geq}) and (\ref{EMeq}) are
\begin{eqnarray}
f(r) &=&\frac{r^2}{2(n-2)(n-3)\alpha} \left
\{1-\sqrt{1-4(n-2)(n-3)\alpha \left( \frac 1{l^2}-\frac
m{r^n}+\frac{q^2}{r^{2n-2}}\right) }\right\} ,
\label{Fg2} \\
A_\mu &=&\frac{(n-1)}{(2n-4)}\frac q{r^{n-2}}((\Xi \delta _\mu
^0-a_i\delta _\mu ^i),\hspace{1.0cm}\text{(no sum on }i\text{).}
\label{pot}
\end{eqnarray}

\section{Physical Properties of the solutions \label{Prop}}

As in the case of rotating black hole solutions of Einstein's
gravity, the above metric given by Eqs. (\ref{met2})-(\ref{pot})
has two types of Killing and event horizons. The Killing horizon
is a null surface whose null generators are tangent to a Killing
field. It is proved that a stationary black hole event horizon
should be a Killing horizon in the four-dimensional Einstein
gravity \cite{Haw1}. This proof can not obviously be generalized
to higher order gravity, but the result is true for all the known
static solutions. Although our solution is not static, the Killing
vector,
\begin{equation}
\chi =\partial _t+{\sum_i^k}\Omega _i\partial _{\phi _i},
\label{Kil}
\end{equation}
is the null generator of the event horizon. The event horizon
define by the solution of $g^{rr}=f(r)=0$. The metric of Eqs.
(\ref{met2})-(\ref{pot}) has two inner and outer event horizons
located at $r_{-}$ and $r_{+}$, provided the mass parameter $m$ is
greater than $m_{\mathrm{crit}}$ given as
\begin{equation}
m_{\mathrm{crit}}=2\left( \frac{n-1}{n-2}\right) \left( \frac
n{n-2}\right) ^{-\frac{n-2}{2(n-1)}}l^{-\frac{n-2}{n-1}}q^{\frac
n{n-1}} .\label{mcrit}
\end{equation}
In the case that
$m=m_{\mathrm{crit}}$, we will have an extreme black brane.

Analytical continuation of the Lorentzian metric by $t\rightarrow i\tau $
and $a_i\rightarrow ia_i$ yields the Euclidean section, whose regularity at $%
r=r_{+}$ requires that we should identify $\tau \sim \tau +\beta _{+}$ and $%
\phi_i \sim \phi_i +i\beta _{+}\Omega _{i}$, where $\beta _{+}$
and $\Omega _{i}$'s are the inverse Hawking temperature and the
angular velocities of the outer event horizon. One obtains
\begin{eqnarray}
\beta _{+} &=&\frac{4\pi \Xi }{f^{\prime }(r_{+})}=\frac{4\pi \Xi
l^2r_{+}^{2n-3}}{nr_{+}^{2n-2}-(n-2)l^2q^2},  \label{bet1} \\
\Omega _i &=&\frac{a_i}{\Xi l^2}.  \label{Om1}
\end{eqnarray}

Usually entropy of the black holes satisfies the so-called area
law of entropy which states that the black hole entropy equals to
one-quarter of horizon area \cite{Beck}. One of the surprising and
impressive feature of this area law of entropy is its
universality. It applies to all kind of black holes and black
strings \cite{Haw2}. However, in higher derivative gravity the
area law of entropy is not satisfied in general \cite{fails}. It
is known that the entropy in the Gauss-Bonnet gravity is
\cite{Wald,Myer}
\begin{equation}
S=\frac 14\int d^{n-1}x\sqrt{\sigma }(1+2\alpha \widetilde{R}),
\label{Entgb}
\end{equation}
where $\widetilde{R}$ is the Ricci scalar for the induced metric $\sigma
_{ab}$ on the $(n-1)$-dimensional horizon. For our case $\widetilde{R}$
vanishes and therefore the entropy is equal to one-quarter of the area of
the horizon as in the case of Einstein gravity, i.e.,
\begin{equation}
S=(2\pi )^k \omega _{n-k-1} \frac{\Xi}4r_{+}^{n-1},\label{Ent}
\end{equation}
where $\omega _{n-1}=(2\pi )^k\omega _{n-k-1}$ is the volume of the
hypersurface at $t=\mathrm{const.}$ and $r=\mathrm{const.}$

The charge of the black brane, $Q$, can be found by calculating
the flux of the electric field at infinity, yielding
\begin{equation}
Q=q \sqrt{\frac{(n-1)(n-2)}2}\left\{ \Xi ^2-\frac{a^2}\alpha \left( 1-\sqrt{1-%
\frac{4(n-2)(n-3)}{l^2}}\right) \right\} ^{1/2}\omega _{n-1}.
\label{chden}
\end{equation}
The electric potential $\Phi $, measured at infinity with respect to the
horizon, is defined by \cite{Cal}
\begin{equation}
\Phi =A_\mu \chi ^\mu \left| _{r\rightarrow \infty }-A_\mu \chi ^\mu \right|
_{r=r_{+}},  \label{Pot}
\end{equation}
where $\chi$ is the null generator of the horizon given by Eq.
(\ref{Kil}). One finds
\begin{equation}
\Phi =\sqrt{\frac{(n-1)}{2(n-2)}}\frac q{\Xi (r_{+})^{n-2}}.
\label{Potel}
\end{equation}

\section{Action, angular momentum and Mass of the uncharged solutions}

\label{Act}

Now we turn to the calculation of the action and conserved
quantities of the black brane in $(n+1)$ dimensions. In general
the action given by Eq. (\ref {Actg}) is divergent when evaluated
on the solutions, as is the Hamiltonian, and other associated
conserved quantities. One way of eliminating these divergences is
to cut off the integral at a large $\mathcal{R}$ and subtract the
solution with $m=q=0$:
\begin{eqnarray}
I_{\mathrm{reg}}=&-& \frac 1{16\pi }\{ \int d^{n+1}x\sqrt{-g}\left( R+%
\frac{n(n-1)}{l^2}\right) \nonumber \\
&-& \frac{f(r=\mathcal{R})}{f(r=\mathcal{R},m=q=0)}%
\int d^{n+1}x\sqrt{-\widetilde{g}}\left( \widetilde{R}+\frac{n(n-1)}{l^2}%
\right) \} .  \label{Actreg}
\end{eqnarray}
The factor $f(r=\mathcal{R})/f(r=\mathcal{R},m=q=0)$ in front of
the second integral is chosen such that the proper length of the
circle which corresponds to the period $\beta _{+}$ in the
Euclidean metric at $r=\mathcal{R}$ coincides with each other in
the two solutions \cite{Odin}. Here we consider only the uncharged
solutions, $q=0$. It is a matter of calculation to show that in
the limit of $\mathcal{R}\rightarrow \infty $, the action is
\begin{equation}
I_{\mathrm{reg}}=-\beta _{+0}\frac{\omega _{n-1}}{16\pi }\frac{r_{+}^n}{l^2}%
=-\frac{\Xi \omega _{n-1}}{4n}r_{+}^{n-1},  \label{Acttot}
\end{equation}
where $\beta _{+0}$ denotes the value of $\beta _{+}$ for $q=0$ which is
\begin{equation}
T=\frac 1{\beta _{+0}}=\frac{nr_{+}}{4\pi \Xi l^2}.  \label{Temp}
\end{equation}
Using Eqs. (\ref{Om1}) and (\ref{Temp}) and the definition of
Gibbs potential, $G(T,\Omega _i)=I_{\mathrm{reg}}/\beta _{+}$, one
obtains:
\begin{equation}
G(T,\Omega _i)=-\frac{(4\pi l^2)^{n-1}}{4n^n} \left(\frac
T{\sqrt{1-l^2\sum_i\Omega _i^2}}\right) ^n\omega _{n-1}.
\label{Gib}
\end{equation}
It is worthwhile to mention that $-(\partial G/\partial T)_\Omega
$ gives the entropy, $S$, given in Eq. (\ref{Ent}), as one
expected. Also one may obtain the angular momentum of the black
brane as
\begin{equation}
J_i=-\left( \frac{\partial G}{\partial \Omega _i}\right) _{T,\Omega
_j}=\frac n{16\pi }\Xi ma_i\omega _{n-1},\hspace{1.0cm}(j\neq i).
\label{Ang}
\end{equation}
Now using the fact that the Gibbs function is the Legendre transformation of
the energy $M(S,J)$ with respect to $S$ and $J$, we obtain
\begin{equation}
M(S,J)=G(T,\Omega _i)+TS+\sum_i\Omega _iJ_i=\frac{\omega _{n-1}}{16\pi }%
(n\Xi ^2-1)m.  \label{Mass}
\end{equation}
One may note that the conserved and thermodynamic quantities
obtained in this section are independent of the parameter
$\alpha$, and therefore they are the same as those computed in the
Einstein gravity.
\section{Local Thermal Stability of the Uncharged Solutions \label{Stab}}

We first obtain the mass $M$ as a function of $S,$ and $J$. Using
the expressions (\ref{Ent}), (\ref{Ang}) and (\ref{Mass}) for the
entropy, the angular momentum, and the mass and the fact that
$f(r_{+})=0$, one obtains by simple algebraic manipulation
\begin{equation}
M=\frac{\omega _{n-1}}n\frac{(n\Upsilon -1)}{\sqrt{\Upsilon (\Upsilon -1)}},
\label{Sma1}
\end{equation}
where $\Upsilon $ is the real positive solution of the following
equation:
\begin{equation}
n^{2n-2}(4S)^{2n}(\Upsilon -1)^{n-1}-(256\pi
^2l^2J^2)^{n-1}\Upsilon =0. \label{Sma2}
\end{equation}
It is worthwhile to mention that the thermodynamic quantities
calculated in Sec. \ref{Act} satisfy the first law of
thermodynamics,
\begin{equation}
dM=TdS+\sum_i\Omega _idJ_i.  \label{1law}
\end{equation}
Furthermore, by using Eqs. (\ref{Om1}), (\ref{Temp}) and
(\ref{Mass}), it is possible to obtain the equation of state for
black brane in the Gauss-Bonnet gravity.

The local stability analysis in any ensemble can in principle be
carried out by finding the determinant of the Hessian matrix
$[\partial ^2S/\partial X_i\partial X_j]\equiv
\mathbf{H}_{X_i,X_j}^S$, where $X_i$'s are the thermodynamic
variables of the system. Indeed, the system is locally stable if
the Hessian matrix satisfies $\mathbf{H}_{X_i,X_j}^S\leq 0$
\cite{Cvet}. In our case the entropy $S$ is a function of the mass
and the angular momentum, but the number of the thermodynamic
variables depends on the ensemble which is used. In the canonical
ensemble, the angular momentum is the fixed parameter, and for
this reason the positivity of the thermal capacity $C_J$ is
sufficient to assure the local stability. The thermal capacity
$C_J$ at constant angular momentum is
\begin{equation}
C_J=T\frac{\partial S}{\partial T}=\frac \Xi 4\left( \frac{(n-2)\Xi ^2+1}{%
(n+2)\Xi ^2-n-1}\right) r_{+}^(n-1).  \label{Cap}
\end{equation}
As one can see from Eq. (\ref{Cap}), $C_J$ is positive for all the allowed
values of the metric parameters discussed in Sec. \ref{Prop}, and therefore
the asymptotically AdS rotating black branes in the canonical ensemble is
locally stable.

In the grand-canonical ensemble, the thermodynamic variables are
the mass and the angular momentum. Direct computation of the
elements of the Hessian matrix of $S(M,J)$ with respect to $M$ and
$J$ is a burdensome task, but the stability requirement
$\mathbf{H}_{M,J}^S\leq 0$ may be rephrased as
$\mathbf{H}_{S,J}^M\geq 0$ \cite{Gus}. It is a matter of
calculation to show that the black string is locally stable in the
grand-canonical ensemble, since the determinant of Hessian matrix,
\begin{equation}
\mathbf{H}_{S,J}^M=\frac{16}{l^2\Xi ^4}\left( \frac{16 \pi}{nl\Xi ^2}%
\right) ^{k-1}[(n-2)\Xi ^2+1]^{-1}r_{+}^{2-n(k+2)},  \label{Hes}
\end{equation}
is positive. It is worthwhile to note that $\alpha $ has not
appeared in the thermodynamic quantities computed in this section.
Thus, the uncharged solutions in the Gauss-Bonnet gravity has the
same thermodynamic features as the solutions in the Einstein
gravity. This fact is also true for the case of static spherically
symmetric black holes with zero curvature horizon \cite{Cai}.

\section{Conclusions and Discussions}

We have found a new class of charged rotating solutions in
Einstein theory with the Gauss-Bonnet term and a negative
cosmological constant. These solutions may be interpreted as black
brane solutions with two inner and outer event horizons provided
the mass parameter $m$ is greater than a critical value given by
Eq. (\ref{mcrit}), and an extreme black brane for the critical
value of the mass. We found that the Killing vectors are the null
generators of the event horizon, and therefore the event horizon
is a Killing horizon for the stationary solution of the
Gauss-Bonnet gravity explored in this paper. The physical
properties of the brane such as the temperature, the angular
velocity, the entropy, the electric charge and potential have been
computed.

We also compute the regulated action of the uncharged solutions
and found the Gibbs potential of the black brane in $(n+1)$
dimensions as a function of temperature and angular velocities. We
obtained the entropy of the black brane by the well known formula,
$S=-(\partial G/\partial T)_\Omega $ and found out that it obeys
the area the law of entropy. Of course this is in commensurate
with the entropy formula of the holes in higher derivative gravity
since the curvature of the horizon is zero for our case. Also the
angular momentum and the mass of the brane are computed.

The first law of thermodynamics has been investigated for the
uncharged rotating black brane and found that the thermodynamic
quantities satisfy this law. Also the local stability of the
asymptotically AdS uncharged rotating black brane both in the
canonical and grand-canonical ensemble was investigated through
the use of the determinant of the Hessian matrix of the mass with
respect to its thermodynamic variables. We showed that the black
branes in various dimensions are locally stable for the whole
phase space both in the canonical and grand-canonical ensemble. We
also found that thermodynamic properties of Gauss-Bonnet rotating
black branes are completely the same as those without the
Gauss-Bonnet term, although the two solutions are quite different.
It seems that the thermodynamics of all the black holes/branes
with zero curvature horizon are robust features of all generally
covariant theories of gravity.

The computation of the action and conserved quantities through the
use of AdS/CFT correspondence and also the stability analysis of
the charged rotating black brane will be given elsewhere.

\end{document}